# Status of the STUDIO UV balloon mission and platform

A. Pahler*[a], M. Ångermann[b], J. Barnstedt[c], S. Bougueroua[a], A. Colin[d], L. Conti[c], S. Diebold[c], R. Duffard[§d], M. Emberger[a], L. Hanke[c], C. Kalkuhl[c], N. Kappelmann[c], T. Keilig[a], S. Klinkner[a], A. Krabbe[a], O. Janson[b], M. Lengowski[a], C. Lockowandt[b], P. Maier[a], T. Müller[e], T. Rauch[c], T. Schanz[c], B. Stelzer[c], M. Taheran[a], A. Vaerneus[b], K. Werner[c], J. Wolf[a]

[a]Institute of Space Systems, University of Stuttgart, Pfaffenwaldring 29, 70569 Stuttgart, Germany;
[b]Swedish Space Corporation, Torggatan 15, 17154 Solna, Sweden;
[c]Institute for Astronomy and Astrophysics, University of Tübingen, Sand 1, 72076 Tübingen, Germany;
[d]Instituto de Astrofísica de Andalucía - CSIC, Granada, Spain;
[e]Max-Planck-Institut für extraterrestrische Physik, Giessenbachstraße, 85741 Garching, Germany

## ABSTRACT

Stratospheric balloons offer accessible and affordable platforms for observations in atmosphere-constrained wavelength ranges. At the same time, they can serve as an effective step for technology demonstration towards future space applications of instruments and other hardware.

The Stratospheric UV Demonstrator of an Imaging Observatory (STUDIO) is a balloon-borne platform and mission carrying an imaging micro-channel plate (MCP) detector on a 0.5 m aperture telescope. STUDIO is currently planned to fly during the summer turnaround conditions over Esrange, Sweden, in the 2022 season. For details on the ultraviolet (UV) detector, see the contribution of Conti et al. to this symposium.[1] The scientific goal of the mission is to survey for variable hot compact stars and flaring M-dwarf stars within the galactic plane. At the same time, the mission acts as a demonstrator for a versatile and scalable astronomical balloon platform as well as for the aforementioned MCP instrument. The gondola is designed to allow the use of different instruments or telescopes. Furthermore, it is designed to serve for several, also longer, flights, which are envisioned under the European Stratospheric Balloon Observatory (ESBO) initiative.

In this paper, we present the design and current status of manufacturing and testing of the STUDIO platform. We furthermore present the current plans for the flight and observations from Esrange.

## 1. STATE OF THE ART IN BALLOON-BORNE ASTRONOMY

The idea of using stratospheric balloons to overcome the opacity of Earth's atmosphere for astronomical observations is not new. Historically, the advantages were obvious: spacecraft did not exist or were hardly available and capabilities of airplanes were limited, leaving balloons as the only option to move instruments above most of the atmosphere. Nowadays, the benefits do not seem as clear: both spacecraft and airplanes provide powerful observation platforms and ground-based telescopes invest large efforts into compensating atmospheric influences. However, even in the era of nano- and microsatellites, space observatories are intrinsically expensive and bear operational limitations: development times are long, updates or corrections of the instrumentation are usually not possible after launch, and consumables, such as cryogenic coolant fluids, cannot be refilled or replaced, as seen with the Herschel Space Observatory. Furthermore, rather conservative approaches towards new technologies are used to minimize risks of expensive failure. Ground-based and airborne telescopes, on the other hand, still suffer from fundamental limitations imposed by the atmosphere at certain wavelengths.

On the other hand, technological advances have made balloon-borne telescopes more attractive over the last couple of years. These particularly include more reliable balloons and the opening of long-duration flight routes, allowing flight durations of 30 to 40 days on "conventional" long duration routes and promising 50 to 100 days on ultra long duration routes. Consequently, the last couple of years saw an increase of balloon-astronomy initiatives aiming at more regularly flying missions rather than the more common "experiment"-type of flights. Noteworthy recent examples are the U.S. / Canadian Superpressure Balloon-borne Imaging Telescope (SuperBIT),[2] the JPL-lead Astrophysics Stratospheric Telescope for High Spectral Resolution Observations at Submillimeter-wavelengths (ASTHROS),[3] and the U.S.-lead Balloon-borne Large Aperture Submillimeter Telescope (BLAST) along with its successors.[4]

*andreas.pahler@irs.uni-stuttgart.de; https://www.irs.uni-stuttgart.de/en/; https://orcid.org/0000-0002-2692-8881
§https://orcid.org/0000-0001-5963-5850

The long-term goal of ESBO is to further lower the entry barrier to balloon-based observations by providing an operating institution that offers observing time and instrument space on balloon-based telescopes. More details on the envisioned ESBO infrastructure and its scope can be found in Maier et al. (2020).[5] The STUDIO platform for UV astronomy described in this paper is the first prototype of an ESBO platform.

The ongoing ESBO *Design Study* (ESBO *DS*) builds on the results of the ORISON project (innOvative Research Infrastructure based on Stratospheric balloONs),[6] and therefore represents the second step towards ESBO. Under ESBO *DS*, the full infrastructure, with particular technical focus on far infrared (FIR) observational capabilities, is being conceptually designed. In addition, a prototype UV / visible flight system (STUDIO) is being developed and built to test some of the key technologies identified.

The next section of this publication focusses on the scientific background for STUDIO. Section 3 explains the gondola and its scientific payload in detail, including the 50 cm telescope and the electronics and software on the payload side. This is followed by a description of the tests conducted on components of the payload in section 4. Finally, section 5 explains the plans for the 2022 flight of STUDIO and plans for the future of ESBO *DS*.

## 2. SCIENTIFIC MOTIVATION

### 2.1. Main Science cases

While atmospheric opacity renders astronomical observations in the ultraviolet (UV) wavelength range impossible from ground, the window between 180 nm and 320 nm opens at altitudes above 37 km. Two science cases have been selected to exploit the unique opportunities that this band – covered by the UV channel of STUDIO – provides: a search for variable hot compact stars in the Galactic plane, and the detection of flares from cool dwarf stars.

Hot and compact stars are the rather short-lived end stages of stellar evolution. They comprise the hottest white dwarfs (WDs) and hot subdwarfs. A significant fraction of them shows light variations with periods ranging from seconds to hours. Among them are diverse types of pulsators, which provide important information to improve asteroseismic models. Others are members of ultracompact binaries (e.g. WD+WD pairs) and are strong sources of gravitational wave radiation. Thus, they are crucial calibrators for future gravitational wave observatories as the planned large ESA mission LISA (Laser Interferometer Space Antenna). Hot compact stars have so far been studied predominantly at high Galactic latitudes. Due to their very blue colors they stick out in old stellar populations like the Galactic halo. However, the density of stars at high Galactic latitudes is rather small and those objects are therefore very rare. Due to the 1000 times higher stellar density, the Galactic disc should contain many more of those objects. Searches in the Galactic plane are desirable, but the identification of these faint stars is hampered by the dense, crowded fields. However, in the UV band hot stars are much easier to detect, because their emitted flux is increasing towards the UV, while the flux of most other stars decreases because of their lower temperatures. Surveying the Galactic plane with a UV imaging telescope will uncover many new variable hot stars.

Red dwarf stars (spectral type M) are hydrogen-burning main sequence stars like our Sun, but lighter, cooler, and fainter. A large fraction of the stars in our Milky Way belongs to this group. They emit most of their radiation in the visible and NIR wavelength regions. Their UV and X-ray emission, despite being energetically a minor contribution to the overall radiation budget, ionizes material surrounding the stars and is, therefore, of central interest for the evolution of planets and other circumstellar matter. This high-energy emission is highly dynamic. One characteristic phenomenon are flares, which are stochastic brightness outbursts resulting from reconfigurations of the magnetic field. During such flares, these normally faint stars become much brighter for the duration of minutes. A strong emission line of ionized magnesium (Mg II) at 280 nm, covered by the STUDIO instrument, can carry up to 50 % of the near-UV flux during flares. Up to now, no systematic UV monitoring of such flare stars exists. Consequently, the flare occurrence rate is unknown as well as the flare energy number distribution. Particularly interesting for the study of the physics of flares is their multi-wavelength behavior (time lags, relative energy in different bands). However, only a few simultaneous UV and optical observations exist. STUDIO enables such observations by continued monitoring of prominent objects.

### 2.2. Possible additional science cases

Additionally, we investigate the possibility to contribute observations to planetary and asteroid science within the solar system. So far, two main science drivers were identified: to monitor composition variations and dynamics of the Venus atmosphere and to determine the composition of asteroids. Of particular interest is the possibility to conduct Venus observations in the UV during day-time, without sacrificing observation time for the main science cases. The attitude control system is expected to work during day-time conditions as well.

## 3. STUDIO – THE UV / VISIBLE PROTOTYPE MISSION

STUDIO is currently under production and integration and will undergo further integration and test activities in 2020 and 2021. Mating of the payload, primarily the telescope, and the gondola is foreseen for early 2021, with subsequent full-system tests on ground. The first flight of STUDIO, which is described in more detail in section 5.1, is currently planned for the autumn turnaround conditions flight window over Kiruna, Sweden, in 2022.

### 3.1. Gondola & Coarse Pointing System

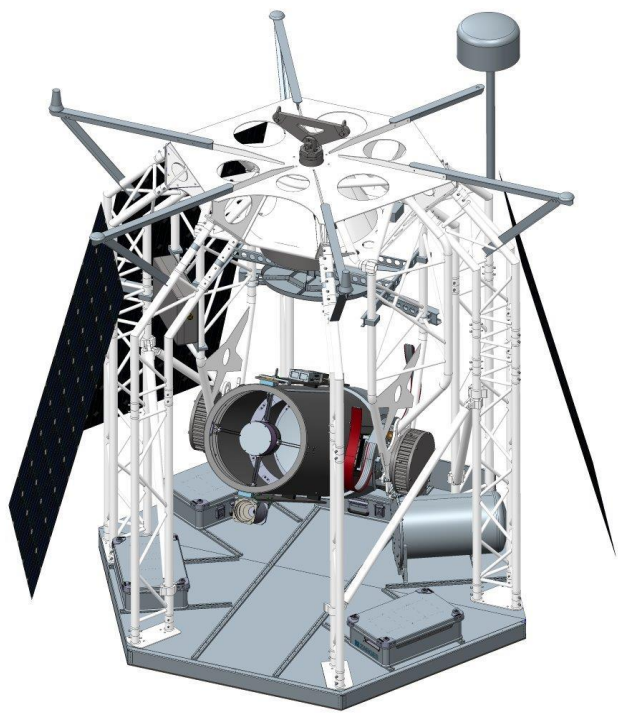

Figure 3.1: STUDIO gondola. One solar panel is not shown for better visibility of the payload. Most electronics and service systems are located on the gondola floor.

The gondola / bus supports the payload with all essential service systems. The mechanical structure is designed from COTS aluminum tube structures in combination with custom-made tubing structures and holds the payload and all the subsystems. The idea is to have a flexible design that can be easily modified for different payload requirements and sizes, and that can also conveniently be disassembled in the field for fast and easy recovery. The structure is covered with white fabric for sun protection that can easily be removed during integration and testing. The telescope and the pointing system are mounted directly in a stiff inner gimbal-like structure, as it can be seen in Figure 3.1. This also makes it possible to test the pointing system with the telescope and the inner structure without the rest of the gondola. When the telescope is placed in its parking position it is covered by a lid that protects it from dust. Most of the subsystems are placed in boxes on the floor of the gondola to achieve a low center of mass for the gondola. This decreases the risk of tipping the gondola at landing and avoiding damages to the telescope and instrument.

The pointing system is a further development of the pointing system[7] developed for the astronomical PoGO balloon missions flown from Esrange in Sweden to northern Canada.[8] It is equipped with an elevation motor directly operating on the telescope and azimuth motors and a fly wheel for turning the complete gondola including the telescope. The pointing system uses multiple sensors, but the most essential ones are three differential GPS antennas placed on booms at the top of the gondola and a star tracker (KU Leuven Star Tracker) placed on the telescope. Further sensors can be added to the system depending on mission requirements and the system can handle heavier telescopes / instruments up to 800 kg.

The other major systems are the power system and the communication system that are based on previous SSC designs from sounding rocket and stratospheric balloon missions.

The power system comprises the solar panels, batteries, and electronic system for power management. It supplies all the systems in the gondola with power, including instrument and pointing system. The batteries and electronics are placed in COTS aluminum boxes for thermal and mechanical protection. Each battery package in the boxes is heated with a foil heater for maximum energy out take during cold periods. The system can easily be scaled according to the mission requirements by increasing or decreasing the number of batteries in the boxes or the number of boxes. In the STUDIO mission, the system is scaled with batteries for covering the nighttime observations and solar panels for daytime charging of batteries. In STUDIO, each main system is supplied with an individual power box. Solar panels are placed on the outside of the gondola for power supply and battery charging. The number and placement of solar panels are modified according to the mission requirements.

The system also comprises electronics and software for controlling the power distribution, monitoring of voltages and currents and switching on and off for the different power outlets. It also comprises housekeeping functions such as temperature, pressure, and a GPS signal that is distributed to the payload.

The communication system that is used for the gondola has two separate communication links. For safety reasons, a separate communication system is implemented for control of the balloon system. For close range communication, when the balloon is within a couple of hundred kilometers from the ground station, a high bandwidth link is used. The range can be extended by using extra ground stations that are placed on the predicted trajectory. For over the horizon or global communication an Iridium-NEXT system will be used with lower bandwidth. This second generation Iridium system gives higher bandwidth and also smaller antennas compared to the previous Iridium systems used in ballooning.

### 3.2. STUDIO Telescope and Optics

STUDIO's optical payload is composed of a 50 cm aperture closed-tube telescope, with the Telescope Instruments Bench (TIP) attached to the back of the telescope. The telescope is a modified Dall Kirkham in an f/13 configuration. It features a 50 cm elliptical primary mirror, a spherical secondary mirror and three lenses for field correction located in the centre hole of the primary mirror. The secondary mirror is moveable via three remotely operable actuators. These are used to move the telescope's M2 mirror with a resolution of ~3 μm, while the telescope's depth of focus is ±30 μm at 180 nm.

The TIP includes two principal instruments:

- An advanced photon-counting, imaging microchannel plate (MCP) detector that performs observations in the UV band (180 nm – 330 nm), described in section 3.3

- A commercial visible light camera PCO.edge 4.2, used mainly as the tracking sensor in a closed-loop fine image stabilization system, which will also serve as a secondary scientific instrument to cover observations in the VIS band (350 nm – 1000 nm)

For details on the camera and its qualification for use in the stratosphere, see section 4.4. Additionally, the TIP houses a fast steering mirror mounted on a commercial Tip / Tilt Platform as part of the image stabilization system to achieve 1 arcsec pointing stability. A dichroic mirror separates the light beam in two channels by reflecting the UV light (180 nm – 330 nm) and transmitting the visible one (330 nm – 1000 nm). The two separated beams are then redirected to the corresponding detector using highly reflecting mirrors. A filter wheel is placed in front of each of the two instruments. Currently it is planned to use one Sloan U filter for the MCP detector, and several for the visible channel. It is possible to integrate additional filters.

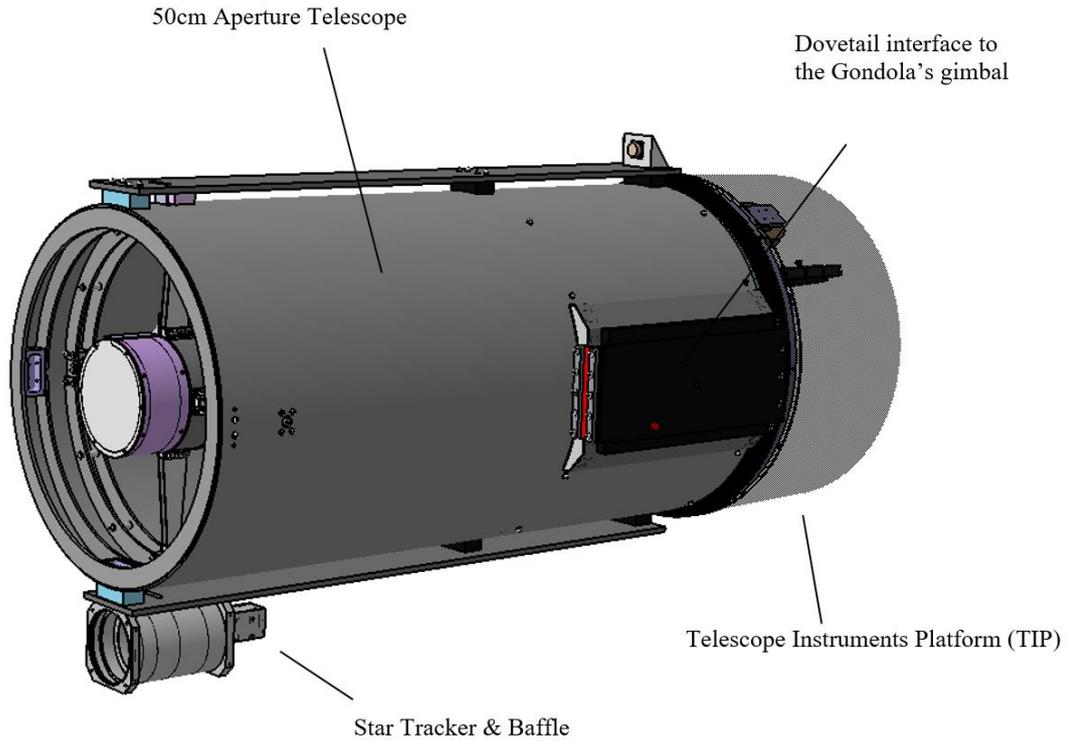

Figure 3.2: STUDIO telescope and TIP assembly

Optical simulations show a promising PSF in the UV channel, with an isolated central peak as presented in Figure 3.3. The corresponding fraction of the enclosed energy as function of the distance from the center is given in Figure 3.4. Eighty percent of the total energy are encircled within a mean diameter of around 17 µm over the wavelength range from 180 nm to 330 nm. This is slightly smaller than the size of one pixel on the UV detector. In the worst case, the PSF is expected to spread over 4 pixels in the UV image plane.

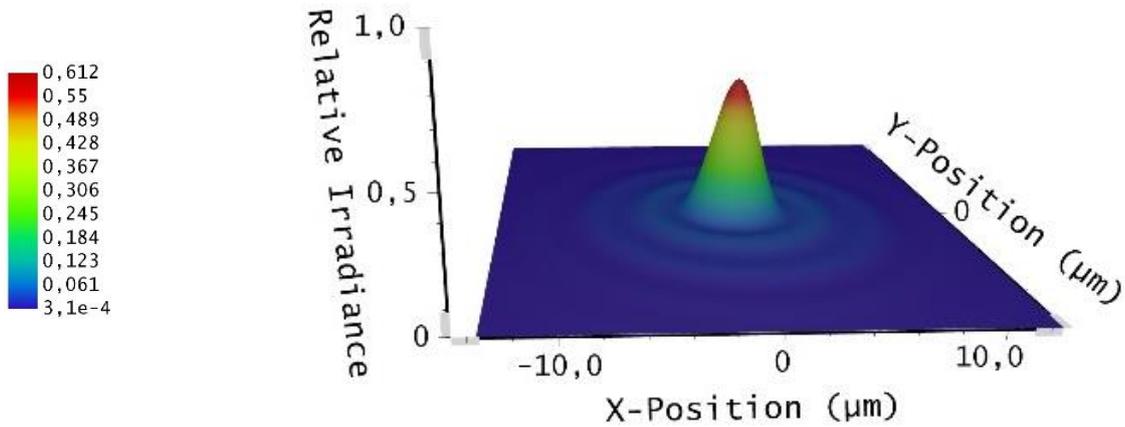

Figure 3.3: Simulated UV PSF over a wavelength range (180-330) nm. Relative irradiance is given as function of the position

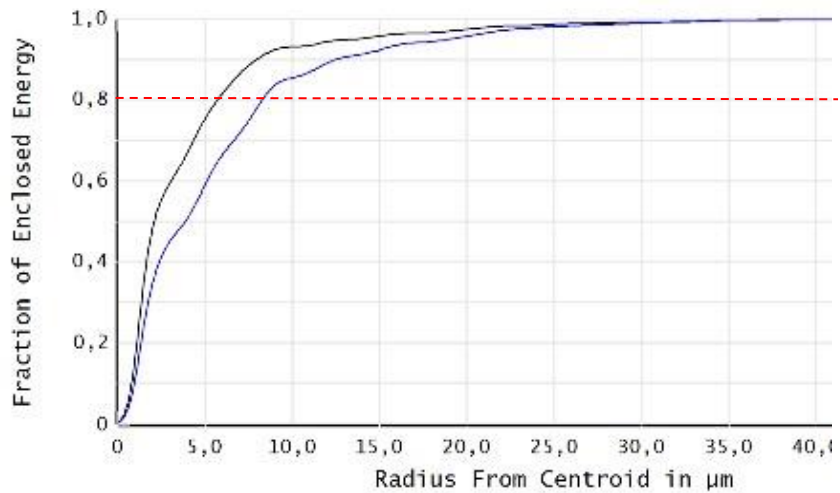

Figure 3.4: The fraction of enclosed energy as function of the radius from the centroid. The dashed line marks 80 % of encircled energy. In black the theoretical, diffraction-limited performance, in blue the simulated performance of the STUDIO system

### 3.3. UV MCP Detector

The UV channel from 330 nm to 120 nm uses a photon-counting microchannel plate (MCP) detector developed at IAAT. Details are presented in Conti et al. (2020).[1] A GaN photocathode converts incoming photons into photoelectrons with an expected quantum efficiency of about 20 %. A detailed description of the development progress can be found in Meyer et al. (2020).[9,10] Each photoelectron is then multiplied into a cloud of electrons in an MCP stack. The center of the electron cloud is correlated with the position of the incident UV photon with a spatial accuracy of about 20 µm. To determine the center of mass of the electron cloud, a cross-strip anode collects the charge on 128 individual strips (64 in x-direction and 64 in y-direction). A preamplifier chip sends the charge information to the front-end electronics (FEE). The photon data are either integrated to an image with 2k × 2k pixels or each detected photon event is transmitted to the instrument computer with its position and a time stamp (photon-by-photon mode). The detector aims to process up to 300,000 events/s in total and 40 to 80 counts/s per pixel, without readout noise and only $2.5 \cdot 10^{-4}$ dark counts/s per pixel at a resolution of 2k × 2k pixels. For a $mag_V$ 19 star (B-V = -0.33) we expect 3 counts/s on the detector and a signal to noise ratio (SNR) of 13 within 1 minute of observation time at low background brightness.

FEE and instrument computer are connected via USB 2.0. At high count rate, the data rate is about 13.2 Mbit/s in photon-by-photon mode. To minimize the number of interfaces, the entire UV detector system uses only one data and one power connection. The power consumption of the detector is less than 18 W in full operation.

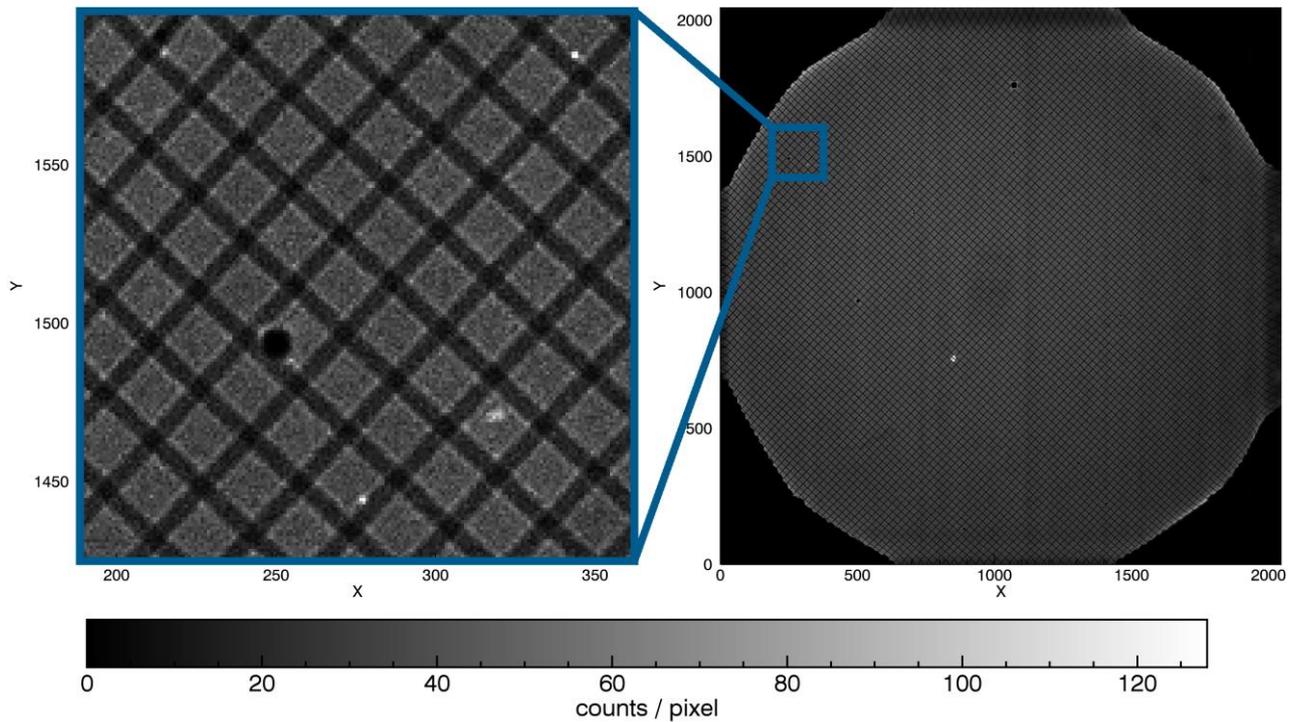

Figure 3.5: Image of a fine grid in front of the detector for test purposes. As can be seen, on most parts of the detector straight lines and even diffraction effects on the grid are visible. Bright spots on the left image originate from defects or irregularities of single MCP channels. For each MCP stack a corresponding look-up table is created to account for these effects.

### 3.4. Mechanical Design of the Payload

The optical bench, on which the TIP components are mounted, is manufactured in-house. It consists of a stiff but lightweight honeycomb sandwich plate made from aluminum and carbon fiber reinforced polymer (CFRP) with a coefficient of thermal expansion equal to that of the telescope backplate.

The instrument mounts are manufactured in-house as well from aluminum 6060. In order to maintain optical alignment of the components over the expected temperature range, the mechanical mounts are designed to be thermally self-compensating. In case of a difference in the thermal expansion between the aluminum and the CFRP, the mounts only allow a movement within the mirror's reflective-surface plane, preventing possible defocus or tilt. Figure 3.6 shows the design of the mirror mounts.

An additional baffle will be placed around the UV light path and will be painted with a UV compatible anti-reflective black paint to minimize light scattering effects. The mirror mounts will be painted with the same paint.

All instruments are kept from overheating using a passive thermal control system. The combination of heat straps and use of the aluminum mounts as heat sinks has shown promising results in thermal simulations. Additionally, temperature sensors and contingency heaters are foreseen for all critical systems. They will allow keeping the devices within their operational temperature range, especially during nighttime when temperatures are expected to drop.

Once all components are installed and aligned, the TIP is closed using a CFRP enclosure. Subsequently, it will be sealed to protect the optics from possible contamination. Contamination protection is supported by a small overpressure (~15 mbar) with dry nitrogen in the optics compartment during ascent.

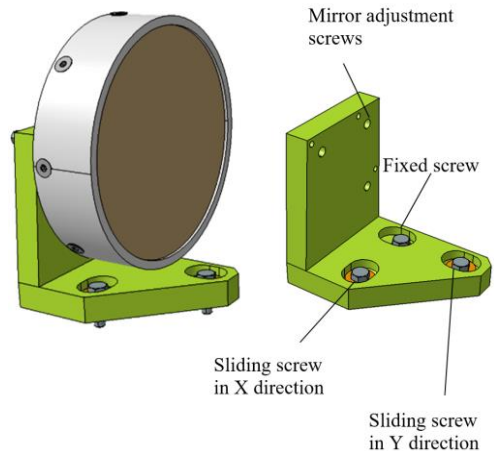

Figure 3.6: Mirror mount

### 3.5. STUDIO Payload Electronics

Commercial off-the-shelf (COTS) components make up a large part of the STUDIO payload. The use of COTS components allowed for major reductions in both cost and development time, since they are more affordable and more easily available, as compared to space-grade components. Furthermore, export regulations do not apply to COTS components, which simplifies the procurement process significantly.

However, the advantages of COTS components are paid with the price of lower reliability and the inability to withstand harsh environmental conditions over a longer time. Although this might seem like a no-go for space missions, COTS components have been used for satellite missions, especially in CubeSats and various University satellites.[11] This suggests that COTS components may also be usable under stratospheric conditions, since the environment is less harsh than in outer space. Nevertheless, components need to be qualified for the use on scientific ballooning missions. If a certain component is found to be inadequate to the environmental conditions, it needs to be modified. If this is not possible, appropriate protection measures from harsh environmental conditions need to be taken.

During the development of the STUDIO payload electronics, a mixture of these approaches was taken, dependent on the capabilities and limitation of individual components. The main payload computer and its close peripherals were enclosed in a pressure housing, in order to protect them from the harsh vacuum environment and to help the thermal management. For parts on the telescope or the optical bench, usage of a pressure-tight vessel was not possible due to space and mass constraints. Therefore, the visible light instrument had to be qualified for use under stratospheric conditions, as described in section 4.4. The same applied for the filter wheels, although some mechanical parts and electronics had to be modified to withstand the extreme environmental conditions. The modifications to the filter wheels are described in more detail in in section 4.5.

### 3.6. Software / FLP Framework

STUDIO implements an architecture that strictly divides hard- and software into gondola / bus and payload, as shown in Figure 3.7. For the bus components and software, mostly components with excessive flight heritage at SSC are used, as also mentioned in section 3.1.

The payload on-board software for STUDIO, on the other hand, is based on a software framework for spacecraft and implements key functionalities and architecture choices used in spacecraft software. Specifically, the Flight Software Framework that was developed for the Flying Laptop satellite is used. It is a component-based software with a core framework that interfaces with different devices and handles inter-process communication between these components.

The framework is platform and operating system independent. ESBO *DS* will use Linux as on-board operating system on an industrial single-board computer, since there are no hard real-time requirements for the prototype mission. Advantages of using this framework include the robustness of the software framework, the clear distribution and hierarchy of functionalities and handling of operating modes and procedures. Moreover, the framework provides the possibility to easily, yet systematically, add and remove components, e.g. for different instruments. In addition, as the core framework was mainly developed for space missions, it uses the ECSS Packet Utilization Standard (PUS) for telecommands and telemetry communication with the payload. For STUDIO, we use part of the RAMSES ground / mission control system to interface to the payload, but the architecture generally allows the flexible use of any other mission control system that implements PUS.

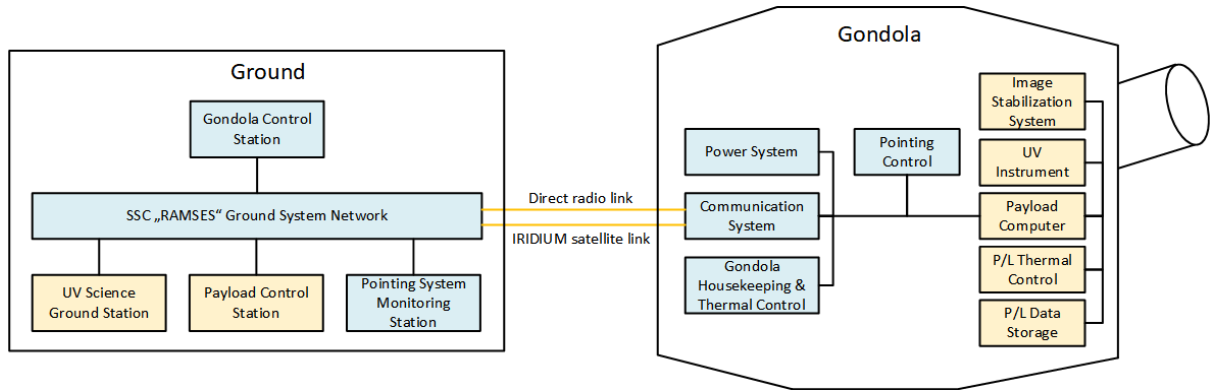

Figure 3.7: STUDIO system architecture. Bus (gondola) components in blue, payload compoents in yellow

For further information on the Flight Software Framework, we refer to Baetz et al. (2017),[12] for further details on its implementation on STUDIO, to Maier et al.[5]

## 4. COMPONENT TESTS FOR STUDIO

### 4.1. Environmental Conditions

When it comes to environmental conditions, the stratosphere is often compared with the space environment, for which extensive literature and standards for testing are available.[13] However, typical balloon flight altitudes of 30 – 40 km still provide a considerable amount of remaining atmosphere. The difference between the hard vacuum of outer space and the coarse vacuum in the stratosphere is favorable for some technical applications, but proves to be more problematic in other aspects.

Furthermore, balloon launches to the stratosphere impose no significant vibrational loads on equipment. However, mechanical loads cannot be neglected completely. A worst-case scenario launch might include the gondola touching ground after release by the launch vehicle. Additionally, the opening of the parachute during descent and the touch-down in unfavorable areas imposes further dangers for the hardware in the gondola. While this might be negligible for single-use payloads, the scope of ESBO *DS* includes the reusability of as many parts as possible between flights. Therefore, a requirement was defined that the payload should survive shocks up to 10 g.

The thermal environment in the stratosphere can be considered close enough to conditions in space for most applications. Heat transmission is dominated by radiation, although small effects of convection have been observed during previous missions.[14] For STUDIO, the International Standard Atmosphere (ISA) atmospheric model has been used to estimate the ambient temperature.[15] However, since convective heat transfer is greatly reduced, and since the gondola provides a certain level of protection, components are not exposed to temperatures as extreme as suggested by the ISA. Components dissipating larger amounts of electrical power may even overheat if no precautions are taken. Therefore, thermal simulations have been conducted to confirm acceptable temperature levels for all components.

Related to the environmental conditions is the question of reliability. Compared to space missions, scientific ballooning missions can tolerate a lower level of reliability, since the hardware only needs to withstand a few dozen hours of flight. Moreover, testing, maintenance, and repair of components is possible in between flights.

### 4.2. Modification of COTS components

Modifications were made to several COTS components in order to harden them against the harsh conditions in the stratosphere, particularly temperature and vacuum. Subsequent qualification tests ensured the effectiveness of these enhancements. As an example, refer to the description of the filter wheel's modifications and qualification in section 4.6.

Electrolytic capacitors are generally avoided in space applications. This is justified by their limited lifetime and pressure sensitivity of some types. For scientific ballooning, these aspects seem of less importance.[16] Nevertheless, for STUDIO, electrolytic capacitors have been replaced on components exposed to stratospheric conditions. Further research will show if this is actually necessary.

Outgassing is considered less critical than in space applications due to the pressure of the remaining atmosphere. For STUDIO, outgassing is of particular concern in close proximity to the optical surfaces. Low-outgassing components and cables were used where possible with reasonable effort. However, some critical parts are not easily available from low-outgassing materials. One example are the CameraLink cables, which will be wrapped with non-outgassing tape to mitigate outgassing.

### 4.3. Environmental Tests

In order to ensure the reliability of components under stratospheric conditions, it was necessary to conduct environmental tests. Testing so far focused on combined thermal and vacuum tests. A shock test will be conducted at system level towards the end of the assembly phase. The test procedure for thermal vacuum testing was derived from the applicable ECSS standard ECSS-E-ST-10-03C,[13] with tailoring and modifications for the specific conditions in stratospheric ballooning. The lower limit of temperature testing is defined to just above -40 °C, with the driver being the capabilities of the thermal vacuum chamber at the Institute of Space Systems at the University of Stuttgart. This temperature is still higher than the environmental temperature according to the ISA. However, as described in section 4.1, components inside of the gondola will not experience these extremely low temperatures. The thermal simulations for STUDIO suggest that the thermal vacuum chamber is able to cover the temperature range expected in flight.

Until now, thermal vacuum testing has been conducted with the visible light camera (described in the following section), the filter wheels (described in section 4.6) and with samples of reflective paint. The latter test aimed at analyzing outgassing from different paints, and is not described in further detail in this paper, since data analysis is still ongoing.

The use of COTS components, some even unmodified, and the in-house development of components clearly highlighted the need of standardized testing. Therefore, qualification procedures are being developed and will be improved to include experience gained throughout the project. This will also facilitate and accelerate the development of future scientific instruments, since working groups are spared the development of their own qualification procedures.

### 4.4. Camera Thermal Vacuum Tests

The image stabilization system of STUDIO relies on a visible light tracking camera.[5] The same camera will be used to support science cases with multi-wavelength observations, or to serve science cases in the visible domain of the spectrum. Since modifications of cameras for the conditions in the stratosphere proved to be too costly for the budget foreseen within ESBO *DS*, an unmodified commercial camera was considered – and eventually chosen – for use in STUDIO. A PCO.edge 4.2 camera was provided for thermal vacuum tests by the manufacturer and was tested without any internal modifications.

Four pressure cycles to 3 mbar were conducted, and several temperature cycles down to -30 °C interface temperature. The camera suffered no damage, and proved to be able to withstand environmental conditions similar to those in the stratosphere. The camera's CMOS sensor is cooled by a built-in Peltier element to 5 °C. For the tests conducted, the camera version with liquid cooling was used. However, the liquid cooling circuit proved to be difficult to seal. Therefore, a passive cooling system is being developed for in-flight use in STUDIO.

At interface temperatures over 17 °C in vacuum, the camera electronics shut off the Peltier cooler due to excessive heat when the liquid cooling system was not used. No noticeable damage was done to the camera or any of the internal components. At interface temperatures below -10 °C, the temperature of the CMOS sensor dropped lower than 5 °C, since the internal Peltier element is not designed to heat the sensor. For scientific applications where the stability of the sensor temperature is critical, external heaters may be used to help stabilize the sensor temperature.

In order to verify proper operation of the camera under these extreme conditions, images of a test target as well as dark frames were taken. The dark current was confirmed to be independent of pressure and in accordance with datasheet values.

As expected, the dark current decreased with temperature to almost zero at -10 °C. In addition to the thermal vacuum tests, the camera was tested in the laboratory and on-sky, as described in section 4.5. Furthermore, it held up to the promises made in the datasheet with respect to image quality and noise, both in vacuum and under standard conditions. Figure 4.1 shows the camera in the thermal vacuum chamber.

To summarize the above, the PCO.edge 4.2 camera was qualified to withstand conditions as expected during the flights of STUDIO, although some limitations became clear during testing. These limitations mainly concerned the cooling in vacuum without the liquid cooling circuit and will be considered in the thermal design of STUDIO.

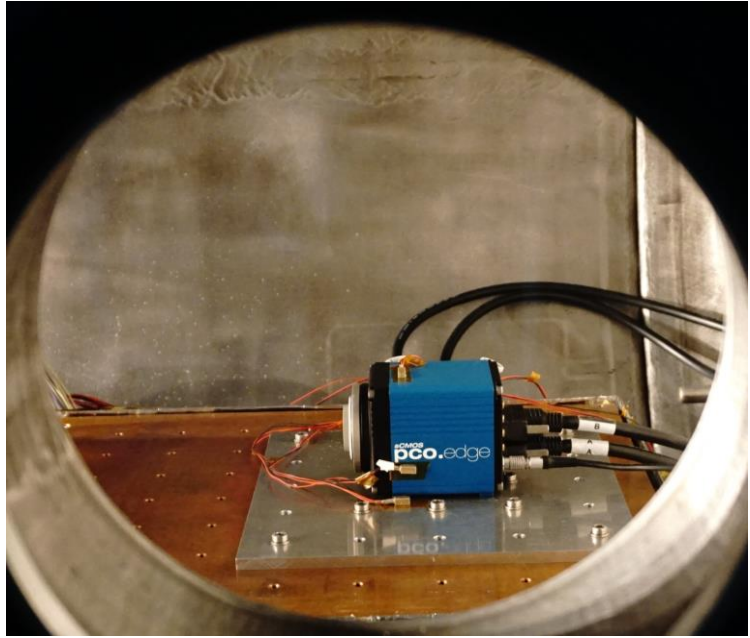

Figure 4.1: PCO.edge 4.2 camera in the thermal vacuum chamber

### 4.5. Camera On-Sky Test

Preliminary tests of the PCO.edge 4.2 camera were performed in laboratory and on-sky to verify its performance and the figures of merits given in the datasheet. Principally, our interest was focused on the frame rate and noise figures; bias, readout and dark current.

The laboratory characterization took place in a class 5 clean room, at ambient temperature and pressure, using an integrating sphere as light source. Most of the parameters measured were slightly better than stated by the manufacturer. A frame rate of 35 FPS was achieved in slow scan of the full 2048 x 2048 pixel frames. Only 0.36 dark electrons/pixel/second were counted at 5 °C sensor temperature. A readout noise of 0.652 e- and a bias of 47.20 e-/pixel were measured.

Following the laboratory tests, the camera was tested on-sky to validate the estimations made on the limiting magnitude of the STUDIO optical assembly. For that purpose a 8" LX200 f/6.3 Meade telescope was used.[17]

The limiting magnitude for different integration times was calculated using the equation:
$$m_{limiting} = m_0 - 2.5 \cdot \log(\sigma \cdot SNR_{min})$$

$m_0$ being the zero magnitude, $\sigma$ the noise and $SNR_{min}$ the minimum SNR – required to be 40 in order to allow a sufficiently accurate determination of the centroid position for the image stabilization system.

The camera is foreseen to be used with the STUDIO 50 cm telescope that has an approximately six times larger collecting area, corresponding to a factor of six of the collected signal. Since the magnitude is an order of 2.5 of the signal, a gain of 2.4 magnitudes is expected. Additionally, the image acquisition at 40 km will provide a better image quality with better seeing.

Results from early simulations have shown that a limiting magnitude of 10.5 (V-band) can be reached at a SNR factor of 40 for an integration time of 10 ms for a combination of the STUDIO telescope and the PCO.edge 4.2 camera. The on-sky tests result with the commercial telescope have been extrapolated to derive the expected performance of the camera using the STUDIO telescope. A limiting magnitude of 10.95 should be achievable at 10 ms of integration time and assuming an SNR factor of 40. These results are therefore in accordance with early estimations.

The Image Stabilization System imposes a limitation of 10 ms on the integration time, which prevents from further improving the limiting magnitude. A way to overcome this issue is to stack multiple images at low integration times. This allows to reach the same sensitivity as that of long exposures, or even better. Experimental results have shown a gain of at least one magnitude achieved by stacking 10 frames of the same integration time.

### 4.6. Filter Wheel Modification and Qualification

In keeping with the goal of incorporating commercially available, off-the shelf components (COTS) in the design of STUDIO, two high-speed filter wheels (HSFW) by Optec, Inc. were chosen for the optical bench. As these devices were not originally meant to be operated in stratospheric environments, a set of modifications needed to be performed to allow usage in such harsh conditions. Subsequently, a rigorous testing regime was completed to ensure that the modifications were successful and adequate. The testing procedure was similar to the one for the visible light camera, described in section 4.4.

The Optec HSFW is a USB-controlled filter wheel that, in the configuration chosen for this project, offers five filter slots with a diameter of 50 mm each. A stepper motor drives the wheel, while a system of magnets and hall sensors provide the built-in microcontroller with information about the current position.

To ready the devices for use in the stratosphere, potentially pressure sensitive electrolytic capacitors were replaced with solid-state ceramic capacitors. Ball bearings in the filter wheels, containing potentially volatile lubricants, were exchanged for non-lubricated plain bearings. To fit the new bearings, special adapters were designed, allowing easy installation and maintainability. O-rings of highly performant methyl-vinyl silicone rubber (MVQ) were installed in place of the original acrylonitrile butadiene rubber (NBR) O-rings. These O-rings are an integral part of the drive system on the HSFW, and their replacement resulted in a significant improvement in low-temperature operability.

Based on ECSS-E-ST-10-03C,[13] a specially tailored test profile was developed using the protoflight approach. With this model, tests are performed directly on the hardware that will be used in the final product, eliminating the need for a separate set of hardware solely used for testing.

A combined total of 17 cycles in a thermal vacuum chamber at the Institute for Space Systems, Stuttgart University, were conducted on the two filter wheels, amounting to over 257 hours. During the test cycles, the filter wheels were examined on their low- and high-temperature functional capabilities as well as their performance in a coarse vacuum environment. Both filter wheels completed their respective test program successfully. Therefore, they are validated for use in stratospheric environments, and on STUDIO.

## 5. STATUS AND FUTURE PLANS

### 5.1. First Flight of STUDIO

The first flight of STUDIO is planned for late summer of 2022 over Esrange Space Center in Northern Sweden. This choice of time allows us to take advantage of the seasonal "turnaround conditions" of the stratospheric winds, which enable payloads to remain practically above the launch site for up to 40 h. Esrange also provides excellent infrastructure, and, compared to other launch sites, easy logistics. Therefore, this setup is very suitable for a first test and science flight. Choosing the turnaround window in August rather than in April / May furthermore allows us to take advantage of the seasonally lower ozone content in the Northern hemisphere, which constitutes the main atmospheric absorbent for UV radiation. The total ozone column over Northern Sweden differs between around 400 Dobson units in late April and around 300 Dobson units in late August.

Figure 5.1 shows the accessible area of the sky during the planned flight. We aim at observing objects in the Galactic plane, to which this flight option provides us with sufficient access.

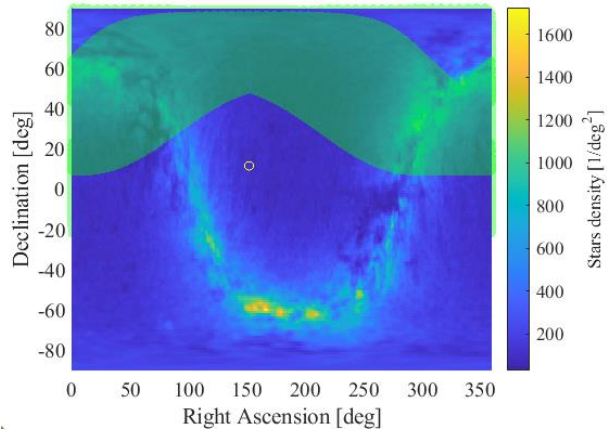

Figure 5.1: Accessible area of the night sky during a summer turnaround conditions flight over Kiruna. Shaded green: accessible area during the flight; background: density plot of all stars in the Tycho-2 catalogue, the milky way is clearly visible as a band with higher density; yellow circle: position of the sun

Table 1 Summary of flight details for the first STUDIO flight

| Duration | 40 h |
| --- | --- |
| Altitude | 37 km |
| Location | Esrange, Sweden |
| Time | August 2022 |
| Gondola mass | 743 kg |
| Balloon size | 600,000 m$^3$ |

In order to lift the 743 kg gondola to 37 km, which are required to reach the necessary atmospheric transmission in the UV (see also Figure 5.2), we are planning to fly with a 600,000 m$^3$ zero pressure helium balloon.

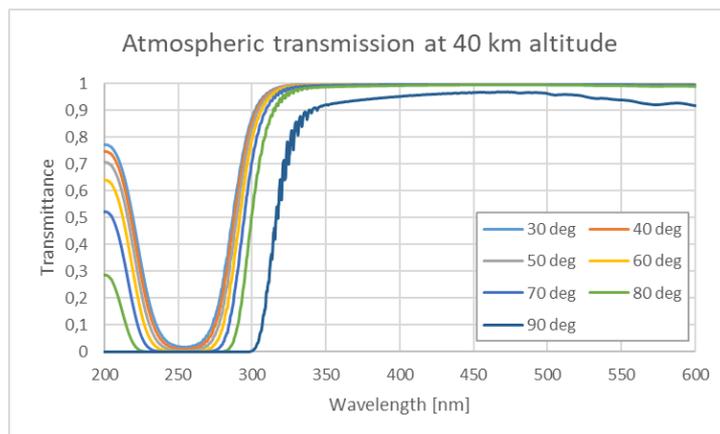

Figure 5.2: Atmospheric transmission in the UV at 40 km altitude for different zenith angles (MODTRAN simulations)

Simulations show that the transmission at 200 nm increases by more than a factor of 2.5 between 30 and 40 km altitude, eventually resulting in a required flight altitude of 37 km.

### 5.2. Future Plans

In line with the goal of the ESBO initiative to provide observations platforms to the community, the STUDIO prototype is prepared to be available for further scientific flights after its maiden flight in 2022. These flights could be carried out either with the same MCP detector, or with a different instrument, for which several interfaces are foreseen. Installation of another detector is possible on the existing optical bench. Furthermore, the optical bench on the back of the telescope can be exchanged, or, if necessary, also the full telescope can be exchanged. Additionally, the gondola is prepared with satellite communication equipment and solar panels, in order to be suitable for longer, e.g. transatlantic flights, offering more observation time per launch.

Besides making the STUDIO prototype available for further flights, near-term future plans particularly include activities to further ease the development and use of astronomical ballooning platforms. This particularly includes the development of standardized test procedures for equipment and the qualification of COTS components for the use on stratospheric missions, as a continuation of activities already started under ESBO *DS*.

Furthermore, one of the main current concerns for reflight of ballooning hardware is the current landing technique, which relies upon unsteered parachute landings. A more controlled and therefore safer landing technique is presented by autonomous, steered landing systems based on steerable parafoils. Such systems are commercially available for military and humanitarian applications. An adaptation and demonstration in a two-stage setup for balloon payloads is planned for upcoming activities.

## 6. SUMMARY

ESBO *DS* is making good progress towards the integration and flight preparation of STUDIO. The first flight of STUDIO is planned to take place in 2022. For this maiden flight, the main science cases are the search for variable hot compact stars and the detection of flares from cool dwarf stars, with further science cases being under consideration. The gondola of STUDIO supports a 50 cm telescope mounted to a gimbal-like structure. A MCP detector will conduct scientific observations in the UV band, while a camera for the visible band is available both for scientific observations and to support the image stabilisation system.

The whole system, including the gondola, is designed to be scalable and flexible. Single components or whole groups of components can be exchanged for future flights. For instance, the detector on the telescope instrument platform could be exchanged, as well as the whole telescope. During the development of STUDIO, space system standards were applied and modified where necessary in order to suit the needs of scientific ballooning.

At the time of publication of this paper, STUDIO is in the production and integration phase and will undergo further integration and test activities in 2020 and 2021.

Future plans include the expansion of the ESBO initiative, including more flights of the STUDIO prototype and to further simplify the use of ballooning platforms for astronomers. This will also include further improvements with regard to reliability and reusability.

## ACKNOWLEDGEMENTS

ESBO *DS* has received funding from the European Union's Horizon 2020 research and innovation program under grant agreement No 777516.

R. Duffard and A. Colin acknowledge financial support from the State Agency for Research of the Spanish MCIU through the "Center of Excellence Severo Ochoa" award to the Instituto de Astrofísica de Andalucía (SEV-2017-0709).

# REFERENCES

[1] L. Conti et al., "Microchannel-Plate Detector Development for Ultraviolet Missions," Proc. SPIE 11444, Space Telescopes and Instrumentation 2020: Ultraviolet to Gamma Ray (11444-17) (2020).

[2] Cho, A., "Cheap balloon-borne telescopes aim to rival space observatories," Science Magazine, https://www.sciencemag.org/news/2020/02/cheap-balloon-borne-telescopes-aim-rival-space-observatories (27 February 2020).

[3] Cofield, C., "NASA Mission Will Study the Cosmos With a Stratospheric Balloon," NASA, https://www.jpl.nasa.gov/news/news.php?feature=7712 (23 July 2020).

[4] Galitzki, N., Ade, P., Angilè. F. E., Ashton, P., Austermann, J., Billings, T. … Williams, P., "Instrumental performance and results from testing of the BLAST-TNG receiver, submillimeter optics, and MKID detector arrays," Proc. SPIE 9914, 99140J (2016).

[5] Maier, P., Ångermann, M., Barnstedt, J., Bougueroua, S., Colin, A. … Wolf, J., "Stratospheric Balloons as a Complement to the Next Generation of Astronomy Missions," 71st Interntational Astronautical Congress, 12-14 October 2020.

[6] Maier, P., Ortiz, J. L., … Alvarez Cuevas, L., "ORISON, a stratospheric project," European Planetary Science Congress, Riga, Latvia 17-22 September 2017.

[7] J.-E. Strömberg, "A Modular Solution for Science Platform Stabilisation," Proc. 20th ESA Symposium on European Rocket and Balloon Programmes and Related Research, Hyère, France (22 – 26 May 2011).

[8] M. Friis, M. Kiss, V. Mikhalev, M. Pearce, and H. Takahashi, "The PoGO+ Balloon-Borne Hard X-ray Polarimetry Mission," Galaxies (6(30)), 1–9 (2018).

[9] K. Meyer et al., "GaN films grown on (0 0 1) and (1 1 0) MgF2 substrates by plasma-assisted molecular beam epitaxy (PA-MBE)," Journal of Crystal Growth (531) (2020).

[10] K. Meyer et al., "X-ray diffraction and secondary ion mass spectrometry investigations of GaN films grown on (0 0 1) and (1 1 0) MgF2 substrates by plasma-assisted molecular beam epitaxy (PA-MBE)," Materials Science in Semiconductor Processing (119) (2020).

[11] Benjamin Schoch, Sebastien Chartier, Ulrich Mohr, Markus Koller, Sabine Klinkner, Ingmar Kallfass, "Towards a Cubesat Mission for a Wideband Data Transmission in E-Band," IEEE Space Hardware and Radio Conference (January 2020).

[12] Bastian Baetz, Ulrich Mohr, Kai Klemich, Nico Bucher, Sabine Klinkner and Jens Eickhoff, "The Flight Software of Flying Laptop: Basis for a reusable Spacecraft Componente Framework," IAA Symposium, Berlin (2017).

[13] European Cooperation for Space Standardization, "ECSS-E-ST-10-03C: Testing," (1 June 2012).

[14] Christopher Geach, "Lessons from PMC-Turbo," Scientific Ballooning Technologies Workshop, Minneapolis (2019).

[15] "Standard Atmosphere," International Organization for Standardization (ISO 2533:1975) (2007).

[16] John Hartley, "Designing For The Stratosphere!," Scientific Ballooning Technologies Workshop, Minneapolis (2019).

[17] Meade Instruments Corporation, "Instruction Manual 8", 10", 12" and 16" LX200GPS Schmidt-Cassegrain Telescopes,", http://www.ruf.rice.edu/~ruco/manuals/LX200GPS%20Manual.pdf.